\documentclass[twocolumn,english,aps,prb,twocolum,superscriptaddress,natbib,bibnotes,amsmath,amssymb,floatfix,groupedaddress,footinbib]{revtex4-2}

\usepackage[colorlinks=true,citecolor=blue,linkcolor=magenta]{hyperref}

\usepackage[markup=nocolor, authormarkupposition=left]{changes} 

\usepackage{soul}
\usepackage[utf8]{inputenc}
\usepackage[english]{babel}
\usepackage{amsmath,amsfonts,amssymb}
\usepackage[T1]{fontenc}
\usepackage{url}

\usepackage{amsmath}
\usepackage{siunitx}
\usepackage[version=4]{mhchem}
\usepackage{amsfonts}
\usepackage{amssymb}

\usepackage{epstopdf}
\usepackage{graphicx}
\graphicspath{{./Figures/}}
\usepackage{gensymb}
\usepackage{threeparttable}

\usepackage{changes}
\DeclareUnicodeCharacter{2212}{-}
\begin{document}

\title[Article Title]{Efficient and Adaptive Reconfiguration of Light Structure in Optical Fibers with Programmable Silicon Photonics}


\author{Wu Zhou}
\author{Zengqi Chen}
\author{Kaihang Lu}
\author{Hao Chen}
\author{Mingyuan Zhang}
\author{Wenzhang Tian}
\author{Yeyu Tong}
\email{yeyutong@hkust-gz.edu.cn}
\affiliation{Microelectronic Thrust, The Hong Kong University of Science and Technology (Guangzhou), 511453, Guangzhou, Guangdong, PR China}

\maketitle

\noindent\textbf{\noindent The demand for structured light with a reconfigurable spatial and polarization distribution has been increasing across a wide range of fundamental and advanced photonics applications, including microscopy, imaging, sensing, communications, and quantum information processing. Nevertheless, the unique challenge in manipulating light structure after optical fiber transmission is the necessity to dynamically address the inherent unknown fiber transmission matrix, which can be affected by factors like variations in the fiber stress and inter-modal coupling. In this study, we demonstrated that the beam structure at the fiber end including its spatial and polarization distribution can be precisely and adaptively reconfigured by a programmable silicon photonic processor, without prior knowledge of the optical fiber systems and their changes in the transmission matrices. Our demonstrated photonic chip can generate and control the full set of spatial and polarization modes or their superposition in a two-mode few-mode optical fiber. High-quality beam structures can be obtained in experiments. In addition, efficient generation is achieved by our proposed chip-to-fiber emitter while using a complementary metal-oxide-semiconductor compatible fabrication technology. Our findings present a scalable pathway towards achieving a portable and reliable system capable of achieving precise control, efficient emission, and adaptive reconfiguration for structured light in optical fibers.}

\section{Introduction}\label{sec1}

The ability to manipulate structured light with a reconfigurable transverse wavefront in terms of amplitude, phase, and polarization has been in demand for various photonics applications \cite{rubinsztein2016roadmap, forbes2021structured}, such as optical trapping \cite{grier2003revolution, stilgoe2022controlled, garces2003observation}, microscopy \cite{hell2007far, maurer2011spatial}, communications \cite{wang2012terabit, bozinovic2013terabit}, imaging \cite{choi2012scanner, caramazza2019transmission}, and quantum information processing \cite{zheng2023multichip, mair2001entanglement}. However, the advancement of arbitrary beam structure  manipulation in optical fibers with flexible light guidance poses challenges \cite{li2023metafiber, bozinovic2013terabit}. Firstly, unknown inter-modal coupling and polarization rotation can occur in a multimode optical fiber, which may also change over time in an uncontrolled manner due to variations in fiber twisting, bending, or even stress. Such time-dependent black box can cause an unpredictable speckle pattern and polarization distribution at the fiber end face. Meanwhile, it is difficult to incorporate a reconfigurable and precise manipulation of the wavefront into the fiber-based systems. Although spatial light modulators \cite{gibson2004free, maurer2011spatial}, cascaded  multiple phase plates \cite{rademacher2021peta, fontaine2019laguerre}, or metasurfaces \cite{shaltout2019spatiotemporal} have shown impressive performance for beam structure manipulation, integrating them with fiber optics requires complex bulky free-space optics that result in additional optical power loss, posing practical challenges for applications. To avoid manipulating beam structure outside of the optical fibers, metasurfaces have also been investigated and applied to the fiber tips using advanced fabrication techniques such as nanoimprinting lithography or 3D laser nanoprinting to realize an all fiber-integrated wavefront shaping system \cite{kostovski2011sub, li2023metafiber, chen2023metasurface}. However, this approach necessitates additional fabrication or assembly process on the fragile fiber tips and are static in nature. Active metasurfaces would be needed to serve for different beam shaping scenarios \cite{shaltout2019spatiotemporal, panuski2022full}.

\begin{figure*}[t]
    \centering
    \includegraphics[width=7in]{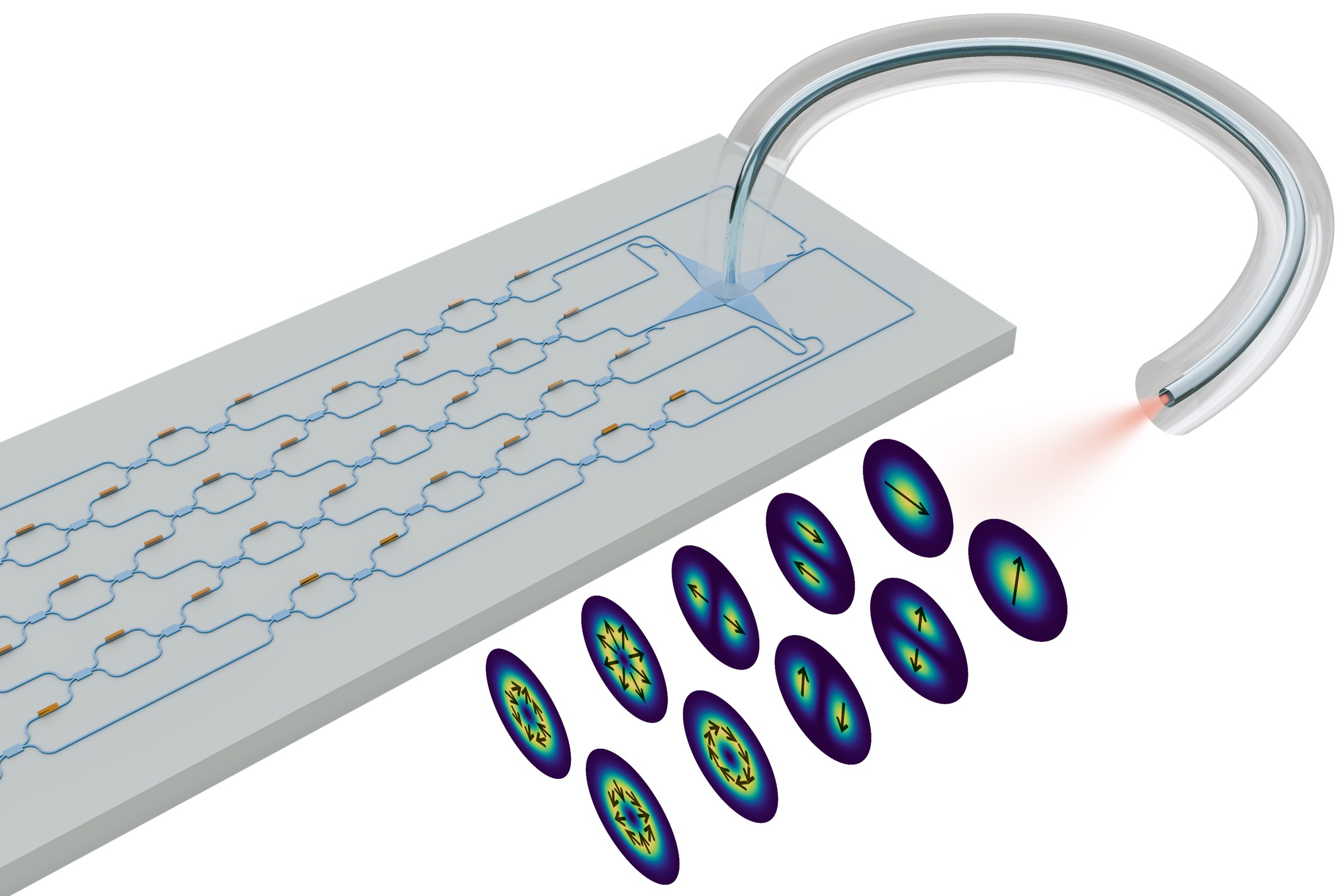}
    \caption{\textbf{Overview of the proposed beam structure shaping system including spatial and polarization distribution at optical fiber far end face.} A 5-meter-long few-mode optical fiber is directly attached to the silicon photonic chip. The integrated photonic processor includes a reconfigurable photonic mesh based on Mach-Zehnder interferometers and an out-of-plane multidimentional chip-to-fiber emitter based on diffraction gratings. By electrically configuring the tunable Mach-Zehnder interferometers inside the photonic processor, the optical field and polarization distribution from the emitter and from the far end face of optical fiber can be tailored in real time accordingly. }
    \label{fig:1}
\end{figure*}

Increasingly in recent years, local generation and manipulation of the free-space structured light via programmable photonic integrated processors have been investigated as they offer an all-integrated and scalable solution on chip \cite{milanizadeh2022separating, butow2024generating, wu2023chip, zhong2024gigahertz, wang2024integrated}. This can largely be attributed to the increasing maturity and rapid advancements of the integrated optics technologies\cite{bogaerts2020programmable}. Silicon photonics platform is currently gaining the most momentum, mainly due to its compatibility with complementary metal-oxide-semiconductor (CMOS) fabrication infrastructures and its capacity to achieve strong optical confinement through ultra-high refractive index contrast \cite{shekhar2024roadmapping}. Precise and lossless control of the on-chip light can be achieved by Mach-Zehnder interferometers or micro-ring resonators based optical mesh within tens of microseconds \cite{miller2013self, ribeiro2016demonstration, carolan2015universal, yi2023reconfigurable}, leading to recent demonstrations in computing accelerators 
 \cite{shen2017deep}, communications 
 \cite{annoni2017unscrambling, lu2024empowering, seyedinnavadeh2024determining}, and quantum information processing \cite{zheng2023multichip}. 
By further incorporating advanced on chip emitters \cite{miller2020analyzing, cai2012integrated, zhou2019ultra, xie2018ultra, liu2018direct}, precise and reconfigurable beam shaping is achievable \cite{milanizadeh2022separating, butow2024generating, wu2023chip}.
Additionally, other advanced photonics functions such as the on-chip laser source may also be incorporated in the future \cite{wan20201, wei2023monolithic}, which holds great promise for the development of a portable and robust structured light generation system. However, existing photonic integration solutions have predominantly focused on manipulating light structure in free space, while less on tailoring the wavefront in optical fibers with low loss and flexible light guiding capabilities \cite{lin2020reconfigurable, yi2023dynamic}. The unique challenges include the need for adaptive structured light control against the unknown fiber transmission matrix \cite{florentin2017shaping, dorrah2022tunable}, the necessity for spatial and polarization distribution manipulation \cite{butow2024generating}, and the lack of efficient chip-to-fiber emitter \cite{tong2019efficient, watanabe2020coherent,zhou2024ultra}. More than that, a rigorous analysis of the beam-shaping method and  quantification of beam quality are essential for verifying the effectiveness of photonic chip solution in achieving accurate and arbitrary wavefront shaping within optical fibers.

Here we demonstrate a programmable silicon photonic integrated processor for local generation and adaptive manipulation of light structure at the other end face of the attached few-mode optical fiber. Arbitrary guided
linearly polarized (LP) beams and their polarization distribution can be precisely reconfigured, despite the existence of unknown and changed optical fiber transmission matrix. The structured light acquired in our experiment demonstrates a superior beam quality, as evidenced by the calculated overlap integrals with the target pattern. To address the issue of limited transmission efficiency between the optical fiber and the photonic chip, we propose and validate a novel multidimensional chip-to-fiber emitter, which exhibits enhanced emission efficiency while maintaining wavefront integrity for all supported spatial and polarization channels. Our demonstrated photonic chip is fabricated by commerical photonics foundries with a CMOS-compatible fabrication process, making it suitable for potential large-scale production in the future. Our results provide a scalable approach towards developing an all-integrated system on chip that enables precise control, efficient emission, and adaptive reconfiguration of light structure within optical fibers.

\section{Results}\label{sec2}

\subsection{Operation Principle}\label{subsec21}

The proposed reconfigurable beam-shaping system for optical fibers with photonic integrated circuits is illustrated in Figure \ref{fig:1}. The system is designed to operate at a wavelength of around 1550 nm within the C band. A flexible optical fiber with a length of about five meters is directly attached to the silicon photonic chip, which will be electrically configured for local and adaptive manipulation of the output field pattern at the other fiber end face. The photonic chip is composed of a reconfiguable optical mesh based on Mach-Zehnder interferometers and a multidimensional chip-to-fiber emitter. The proposed mutidimensional emitter is based on out-of-plane two-dimensional (2D) diffraction gratings, which are different from the chip-to-free space emitter employing a grating coupler array where each of the grating couplers has a fixed emitted field and polarization distribution after fabrication \cite{milanizadeh2022separating, butow2024generating}. Our proposed multidimensional emitter can support launching of the full set of fiber LP modes in the two orthogonal polarizations into a two-mode few-mode fiber. To achieve a precise manipulation over the spatial and polarization distribution at the output fiber end face, adaptive wavefront tailoring is essential. As random polarization rotation or inter-modal coupling can occur due to birefringence caused by fiber stress variations, fiber bending, or misalignment of fiber splices. In our design, we utilize a reconfigurable integrated optical mesh to adjust both the amplitude and phase of the coherent light directed into the emitter. This enables us to manipulate the wavefront of the optical field that is coupled into the attached optical fiber. Through precise and local configuration of the integrated optical mesh to invert the transmission matrix of the optical fiber, adaptive wavefront control at the fiber end face can be achieved. To validate our proposed method, a few essential aspects should be further investigated in our design and experiments. First of all, the feasibility of achieving an efficient and multidimensional chip-to-fiber emitter needs to be assessed. Secondly, the dynamic wavefront control using the on-chip photonic mesh in the presence of unknown and changed fiber transmission matrix should be examined. Most importantly, beam structure from the on-chip emitter and from the far end of the fiber should be collected to assess the similarity between the desired structure with the experimental results.

\subsection{Chip-to-fiber multidimensional emitter design}\label{subsec22}

\begin{figure*}[t]
    \centering
    \includegraphics[width=1\linewidth]{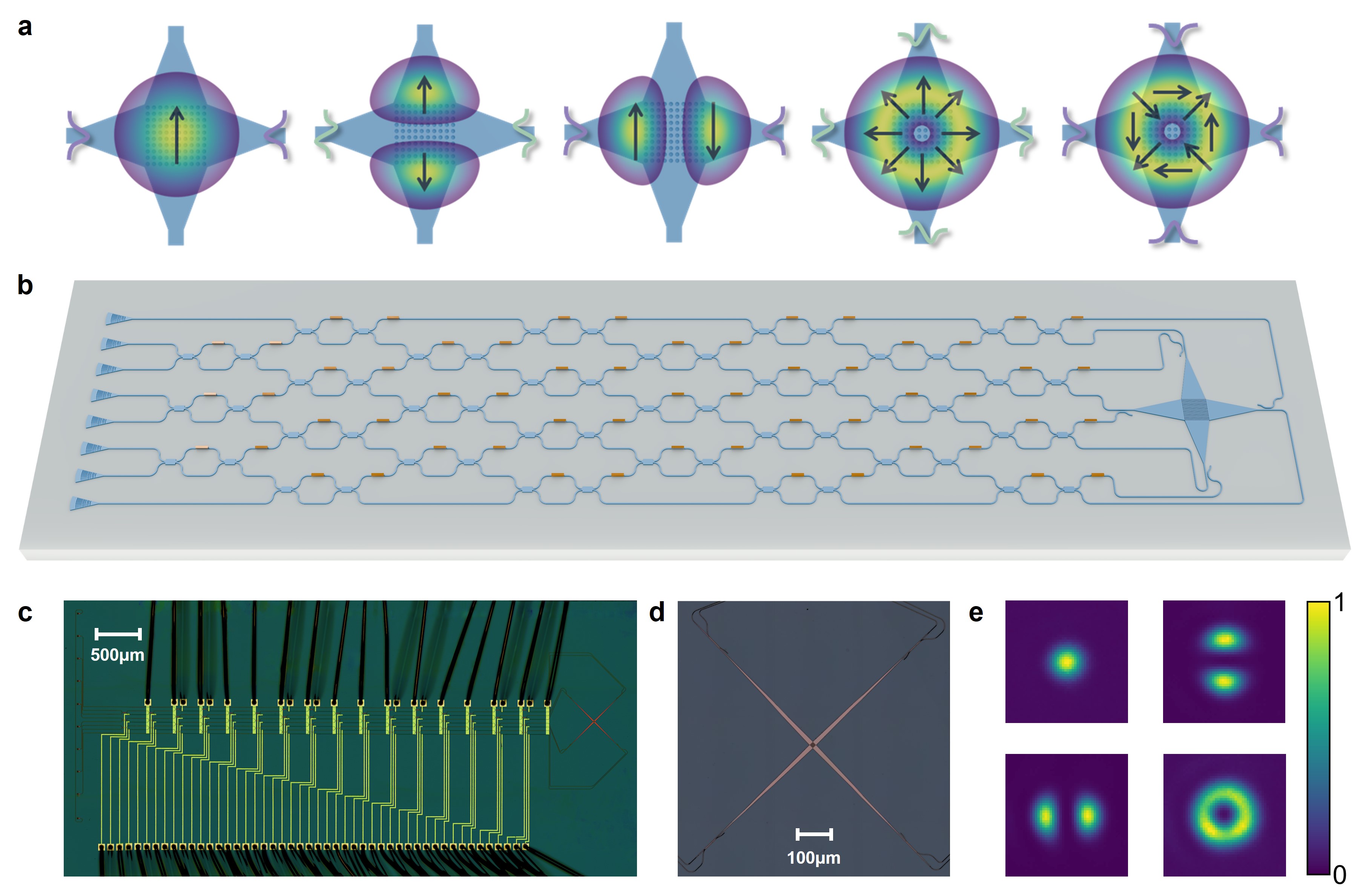}
    \caption{\textbf{Design schematic and microscopy image of the integrated photonic processor.} \textbf{a} Selective launching conditions of various fiber modes and their superposition by the proposed 2D diffraction gratings. The diffracted field distribution is determined by injected waveguide modes and their relative difference in phase and amplitude. \textbf{b} Schematic of integrated photonic processor for fiber wavefront shaping, consisting of input single-mode grating couplers, a photonic mesh based on Mach-Zehnder interforemetors, and a multidimensional emitter chip-to-fiber emitter. \textbf{c} microscopy image of the wire-bonded photonic chip. \textbf{d} Zoom-in view of the multidimensional chip-to-fiber emitter fed with four perpendicular waveguides. \textbf{e} diffracted optical field from a pure multidimensional chip-to-fiber emitter in experiment captured by an infraded camera under various selective launching conditions.}
    \label{fig:2}
\end{figure*}

Our proposed method necessitates a multidimensional chip-to-fiber emitter capable of supporting a wide range of spatial and polarization channels, all while maintaining high coupling efficiency and mode-selective launching capabilities. Furthermore, a CMOS-compatible fabrication process should be employed to enable mass production by photonics foundries in the future. However, the conversion efficiency between the on-chip strip waveguide modes and the higher-order fiber modes is one of the major concern. Previous studies have investigated both in-plane and out-of-plane chip-to-fiber coupling strategies to improve the number of supported spatial and polarization channels and coupling efficiency \cite{koonen2012silicon, tong2019efficient, watanabe2020coherent, zhou2024ultra, xiaolin2024efficient}. In our demonstration, we utilize a 2D diffraction grating for accessing the complete set of spatial and polarization channels allowed in a two-mode few mode optical fiber, including two \( \text{LP}_{\text{01}} \) modes polarized in two perpendicular directions and four degenerate \( \text{LP}_{\text{11}} \) modes, as depicted in Figure S1. On the planar photonic chip, optical signals injected from four perpendicular directions converge on a centrosymmetric square 2D grating region. Two on-chip transverse electric (TE) waveguide modes, including the fundamental TE mode \( \text{TE}_{\text{0}} \) and the first-order \( \text{TE}_{\text{1}} \) mode, are fed into the 2D grating region with active phase and amplitude control. Selective launching of various linearly polarized modes or their superposition can thus be achieved by the multidimensional emitter as illustrated in Figure \ref{fig:2}a. In our design, beam shaping occurs within the same grating region before being coupled into the few-mode fiber attached. When the diffracted beam diameter is designed to be well matched with the mode size supported by optical fiber, a wavefront-manipulable optical signal can be efficiently sent into the few-mode fiber.

Figure \ref{fig:2}b illustrates full schematic of the proposed photonic processor, including the chip-to-fbier emitter, the photonic mesh, and the input single-mode grating coupler array. This configuration is employed due to the utilization of an off-chip laser source in our demonstration. Microscopy image of the wired bonded photonic chip under test is shown in Figure \ref{fig:2}c. A zoom-in view of the multidimensional chip-to-fiber emitter is presented in Figure \ref{fig:2}d. We first characterize the individual chip-to-fiber emitter in experiment. By feeding on-chip waveguide modes with appropriate configurations outlined in Figure \ref{fig:2}a, various fiber modes including their superposition can be selectively launched with diffracted optical field captured by an infrared camera shown in Figure\ref{fig:2}e. The full set of six spatial and polarization channels in the two-mode few-mode fiber can be selectively excited with measured field distribution shown in Figure S2e. After that, one end of the few-mode optical fiber is placed over the chip-to-fiber emitter. By monitoring the output optical power from the attached optical fiber, chip-to-fiber coupling efficiency can be measured with transmission spectrum shown in Figure S2h. The measured peak chip-to-fiber efficiency for \( \text{LP}_{\text{01}} \), \( \text{LP}_{\text{11a}} \), and \( \text{LP}_{\text{11b}} \) modes are -3.5 dB, -6.1 dB, and -4.3 dB, respectively. Other detailed design parameters and experimental results of the multidimensional emitter are included in the Supplementary Information S2. A higher efficiency is desired but becomes difficult as it has already approached the typical performance limit of the most reported 2D grating couplers \cite{marchetti2019coupling}. While subwavelength structures or bottom mirrors have the potential to increase the amount of optical energy diffracted upwards, achieving this would necessitate a very small minimum feature size or involve complex fabrication steps \cite{watanabe20192, luo2018low}. 

\begin{figure*}
    \centering
    \includegraphics[width=0.8\linewidth]{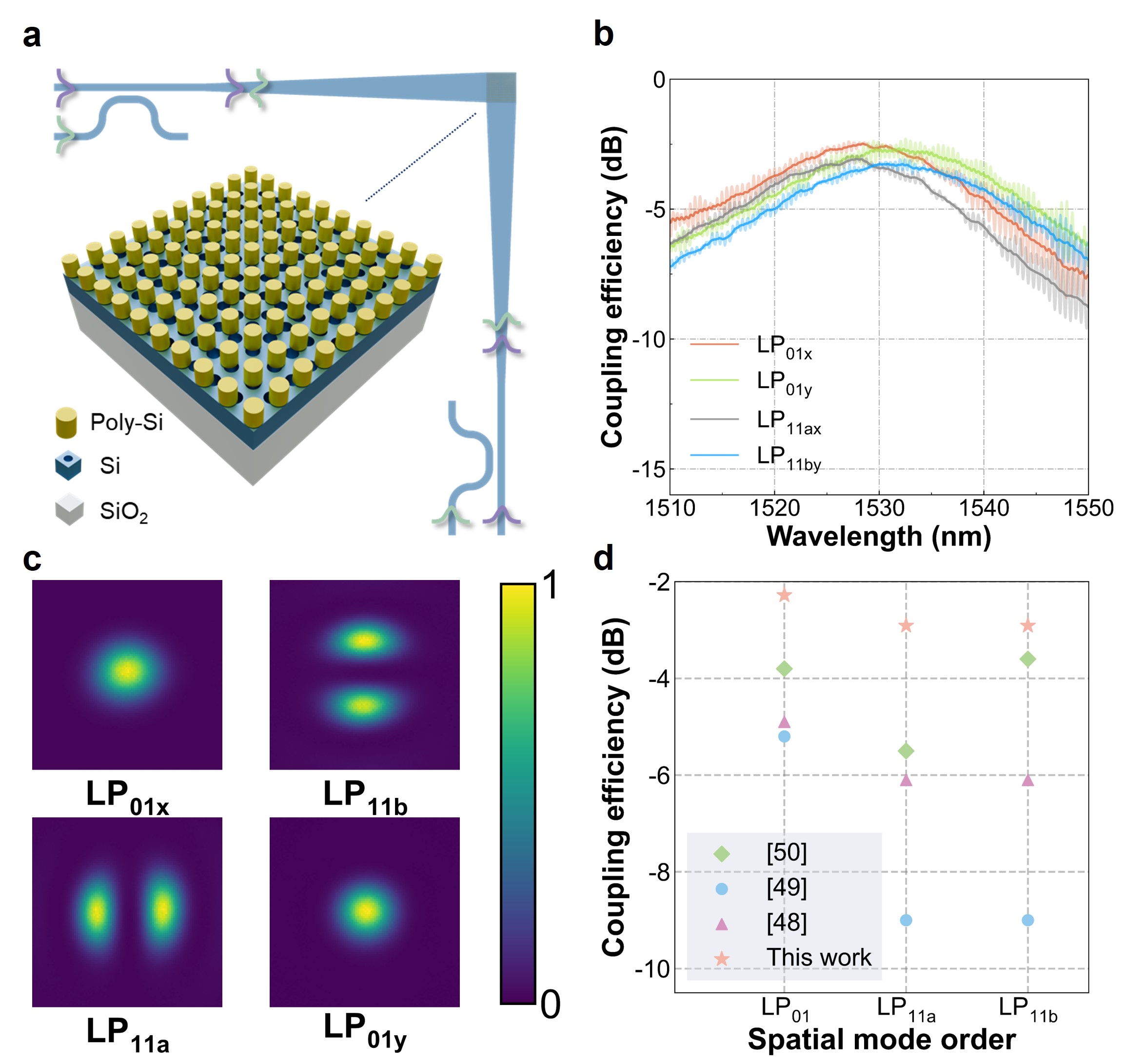}
    \caption{\textbf{Design schematic and experimental results of the polycrystalline silicon (poly-Si) overlay enhanced multidimensional chip-to-fiber emitter.} \textbf{a} Schematic of the emitter based on 2D grating coupler, consisted of 70-nm shallowly etched hole array and shifted 160-nm thick Poly-Si disk array. \textbf{b} experimental coupling efficiencies of different fiber modes including \( \text{LP}_{\text{01x}} \), \( \text{LP}_{\text{01y}} \), \( \text{LP}_{\text{11ax}} \), and \( \text{LP}_{\text{11by}} \). \textbf{c} diffracted optical field captured by an infrared camera from the Poly-Si overlay enhanced multidimensional emitter in experiment under various selective launching conditions. \textbf{d} Comparison of the chip-to-fiber coupling efficiencies with other recent work using diffraction gratings on silicon.}
    \label{fig:3}
\end{figure*}

To solve this issue, in this work, we also propose and demonstrate a new multidimensional 2D grating structure with high coupling efficiency while compatible with the CMOS fabrication process using optical lithography. Compared to the emitter shown in Figure \ref{fig:2}d, the proposed multidimensional emitter is based on the same 220-nm thick SOI wafer with 2$\mu$m-thick buried oxide. 160-nm thick polycrystalline silicon (poly-Si) disks are deposited on top of the previous grating region, as illustrated in Figure \ref{fig:3}a. The designed diameter of each poly-Si disk is 300 nm. By breaking the vertical symmetry of the grating region using optimized front-end-of-line process, the directionality of the grating coupler can be effectively enhanced, thus allowing more optical energy to be coupled upwards. Apart from that, the placement of shallow-etched holes and poly-Si disks are intentionally staggered in our design as illustrated by Figure S3b. This is necessary to precisely control the grating strength in the multimode diffraction region. Hence, a small effective refractive index contrast can be realized to engineer the diffracted beam size and match it with the guided wave in the optical fiber. Detailed design method and simulation results are elaborated in the Supplementary Information S3. It is worthwhile to mention that, unlike the above multidimensional 2D grating emitter fed from four perpendicular directions, the multidimensional emitter enhanced by poly-Si overlay shown in Figure \ref{fig:3}a is only fed from two perpendicular directions. In this case, selective mode launching of \( \text{LP}_{\text{01x}} \), \( \text{LP}_{\text{01y}} \), \( \text{LP}_{\text{11ax}} \), or \( \text{LP}_{\text{11by}} \) can be easily realized by injecting a single \( \text{TE}_{\text{0x}} \), \( \text{TE}_{\text{0y}} \), \( \text{TE}_{\text{1x}} \), or \( \text{TE}_{\text{1y}} \), respectively. No active phase or amplitude control is needed. Due to the reciprocity of light, it can also be used at the opposite end of the optical fiber as a receiver to evaluate the purity of spatial and polarization distribution for the four physical channels, by measuring their optical power transmission in our later experiment. This cannot be achieved solely through the use of a polarization-insensitive infrared camera.

Our proposed new emitter was also fabricated by a commercial silicon photonics foundry. The measured chip-to-fiber coupling efficiency spectra for the four spatial and polarization channels are shown in Figure \ref{fig:3}b. A peak coupling efficiency of -2.28 dB and -2.39dB is achieved for \( \text{LP}_{\text{01x}} \) and \( \text{LP}_{\text{01y}} \), while -2.91 dB and -3.16 dB is attained for \( \text{LP}_{\text{11ax}} \) and \( \text{LP}_{\text{11by}} \). The corresponding emitted field distribution for various linearly polarized modes are captured by an infrared camera as depicted in Figure \ref{fig:3}c, which can confirm the improved efficiency while maintaining the transverse wavefront integrity of both the fundamental and higher-order modes. Figure \ref{fig:3}d summarizes the experimental coupling efficiencies of the emitters reported in recent works on silicon photonics platform for few-mode optical fibers. Our proposed emitter can effectively increase the coupling efficiency for various spatial and polarization modes, while compatible with the current silicon photonic fabrication process using optical lithography. In the future, optical signals can also be injected from four perpendicular directions to enable efficient launching of the full set of six orthogonal spatial and polarization channels in the two-mode few-mode fiber shown in Figure 2d.

\subsection{Photonic mesh design and configuration}\label{subsec23}

As the structure of diffracted beam via multidimensional emitter depends on the relative amplitude and phase of the eight injected on-chip waveguide modes shown in Figure \ref{fig:2}a, wavefront shaping can be allowed by a precise control of on-chip light through a reconfigurable photonic mesh before being injected into the multidimensional emitter. In our demonstration, the proposed photonic mesh depicted in Figure \ref{fig:2}b is based on an 8-degree unitary transformation matrix $U_{8}$ consisted of 28 integrated Mach-Zehnder interferometers by taking rectangular decomposition, where each of the Mach-Zehnder interferometers acts as a 2$\times$2 unitary block $U_{2}$. Two heater-based thermo-optic phase shifters are employed in each unitary block to adjust the power splitting ratio and the relative phase delay. At the output of the photonic mesh, four of the on-chip fundamental \( \text{TE}_{0} \) modes are multiplexed into the \( \text{TE}_{1} \) modes by evanescent coupling using four asymmetrical adiabatic directional couplers, as depicted by Figure S2c. Hence, four input ports of the multidimensional emitter can all be accessed, where each port contains amplitude and phase controllable \( \text{TE}_{0} \) and \( \text{TE}_{1} \) mode. Detailed design of the on-chip mode multiplexer by asymmetrical adiabatic directional coupler can be found in the Supplementary Information S2. 

Mathematically, the transmission matrix of a two-mode few-mode fiber can be represented by an unknown 6-degree unitary transformation matrix $U_{6}$. As the full set of six orthogonal basis vectors in the two-mode few-mode fiber can all be accessed via our multidimensional emitter as illustrated in the Supplementary Information S2, the transformation matrix of the emitter can be formulated as a  semi-unitary \( M_{6 \times 8} \) shown in Equation S6. The end-to-end transmission matrix of the entire system can be expressed as shown in Equation \ref{equ1}. The on-chip reconfigurable unitary transformation $U_{8}$ can always be decomposed into the product of four matrices: the conjugate transpose of the emitter transmission matrix $M_{6 \times 8}^{\dagger}$, the conjugate transpose of the optical fiber transmission matrix $U_{6}^{\dagger}$, the emitter transmission matrix $M_{6 \times 8}$, and a unitary matrix $U_{8}$. Consequently, when a coherent beam of light is fed into the on-chip processor via an external laser source and one of the input single-mode grating couplers in our experiment, the on-chip photonic mesh $U_{8}$ can be actively programmed to invert the transmission matrix of the optical fiber, and arbitrarily rotate the input vector with a constant modulus. All possible linear combinations of the six orthogonal fiber modes can be achieved at the fiber end adaptively and locally. Detailed mathematics and matrix operation are provided in the Supplementary Information S3 to demonstrate capability of our proposed photonic processor for arbitrary structured light manipulation in the attached two-mode few-mode fiber.

\begin{equation}
T = U_{6} \cdot M_{6 \times 8} \cdot U_{8} = U_{6} \cdot M_{6 \times 8} \cdot M_{6 \times 8}^{\dagger} \cdot U_{6}^{\dagger} \cdot M_{6 \times 8} \cdot U_{8}
\label{equ1}
\end{equation}

To further validate adaptive wavefront control against the existence of unknown and changed polarization rotation and inter-modal coupling in the attached optical fiber, additional experimental results are required to examine the reconfigurability of the photonic processor and to quantify the degree of resemblance between the structure obtained at the fiber end face and the target structure. In our experiment, the Mach-Zehnder interferometers within the wire-bonded photonic processor are controlled by a multichannel programmable power supply. A thermoelectric cooler is employed for thermal stabilization of the integrated photonic processor. We use particle-swarm optimization algorithm \cite{kennedy1995particle} to dynamically control the beam structure emitted from the chip and from the fiber end face, with the feedback from the real-time calculated overlap integrals between the captured field profile by the infrared camera and the target. Details of the structure quality characterization are included in the Supplementary Information S4.

\subsection{Experimental setup and beam-shaping results}\label{subsec24}

\begin{figure*}
    \centering
    \includegraphics[width=0.9\linewidth]{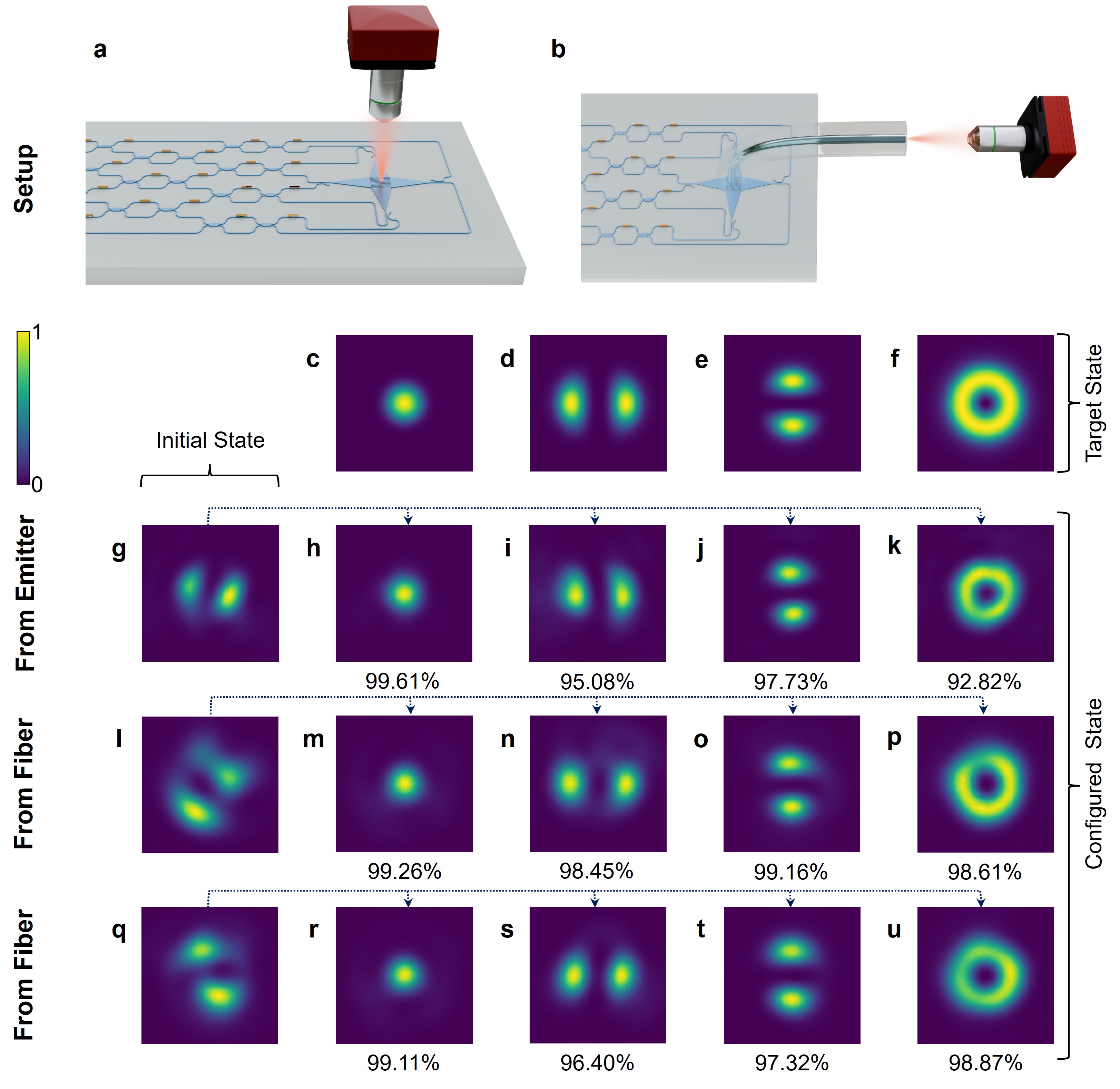}
    \caption{\textbf{Experimental setup and results of the structured light generated from the emitter and fiber end face.} \textbf{a} Experimental setup to measure the beam structure from the multidimensional emitter without optical fiber. \textbf{b} Experimental setup to measure the beam structure from the output end face of the optical fiber attached. \textbf{c-f} Target beam structure obtained in simulation. \textbf{g} random initial beam structure on top of the multidimensional emitter with unconfigured photonic mesh. \textbf{h-k} structured light generated from the emitter after configuring the photonic mesh according to the target pattern. \textbf{l} random initial beam structure at the other end of the optical fiber with unconfigured photonic mesh. \textbf{m-p} structured light generated from the fiber end after configuration of the photonic mesh according to the target pattern. \textbf{q} another random initial beam structure at the other end of the optical fiber with unconfigured photonic mesh and different fiber stress. \textbf{r-u} structure generated from the fiber end after configuration of the photonic mesh.}
    \label{fig:4}
\end{figure*}

The experimental setup is shown in Figure \ref{fig:4}a and \ref{fig:4}b, where a 10x microscope objective and infrared camera are used to image the the emitted structured light from the photonic processor and from the other fiber end face, respectively. Before being launched into the optical fiber, our demonstrated photonic processor can also be utilized for free-space structured light generation. When a coherent beam is sent into the photonic chip by an external tunable laser source using one of the input single-mode grating couplers, the output spatial light from the multidimensional emitter will be random without calibrating the photonic mesh, as depicted by Figure \ref{fig:4}g. Next we can optimize the configuration of the photonic processor according to the real-time calculated overlap integrals between the measured one with the target patterns shown in Figure \ref{fig:4}c-\ref{fig:4}f. All the target structures can be successfully configured as shown in Figure \ref{fig:4}h-\ref{fig:4}k, including \( \text{LP}_{01} \), \( \text{LP}_{\text{11a}} \), \( \text{LP}_{\text{11b}} \), and the superposition of \(\text{LP}_{\text{11a}} \) with \( \text{LP}_{\text{11b}} \). All exhibit overlap integrals with the target structure better than $92\%$. After the above calibration process, all the free-space structured light can be obtained from the emitter quickly using the same optimized configurations without any additional feedback control.

Although free-space structured light from the multidimensional emitter can well resemble the target beam, once being coupled into the attached optical fiber, its wavefront can be quickly distorted due to existence of polarization rotation and inter-modal coupling. Thus, adaptive wavefront shaping is needed which can be achieved by the programmable photonic processor. Similarly, we put the microscope objective and infrared camera at the end face of the attached optical fiber. Figure \ref{fig:4}l shows a completely distorted wavefront at the fiber end face when a pure \( \text{LP}_{11} \) mode is launched into the optical fiber. The phototnic processor is then reconfigured according to the target structure at the fiber output. The obtained beam structures and the corresponding overlap integrals are depicted in Figure \ref{fig:4}m-\ref{fig:4}p, which can demonstrate the reconfigurability of our proposed photonic processor. We also observe that the beam field diameter from the emitter is slightly smaller than that from the optical fiber, resulting in a slight penalty for the higher-order modes in their overlap integrals with the target pattern. This also suggests that an increase in emitter size will be helpful in the future to enhance the mode field overlap and its coupling efficiency. Subsequently, the few-mode optical fiber is deliberately displaced and twisted to exhibit another unknown transmission matrix, as illustrated by the different initial random structure depicted in Figure \ref{fig:4}q. By resetting the configuration of the photonic processor, similar target beam structures shown in Figure \ref{fig:4}r-\ref{fig:4}u can be retrieved at the fiber end, which can demonstrate the adaptive reconfiguration ability of our propose method.

\begin{figure*}[t]
    \centering
    \includegraphics[width=1.0\linewidth]{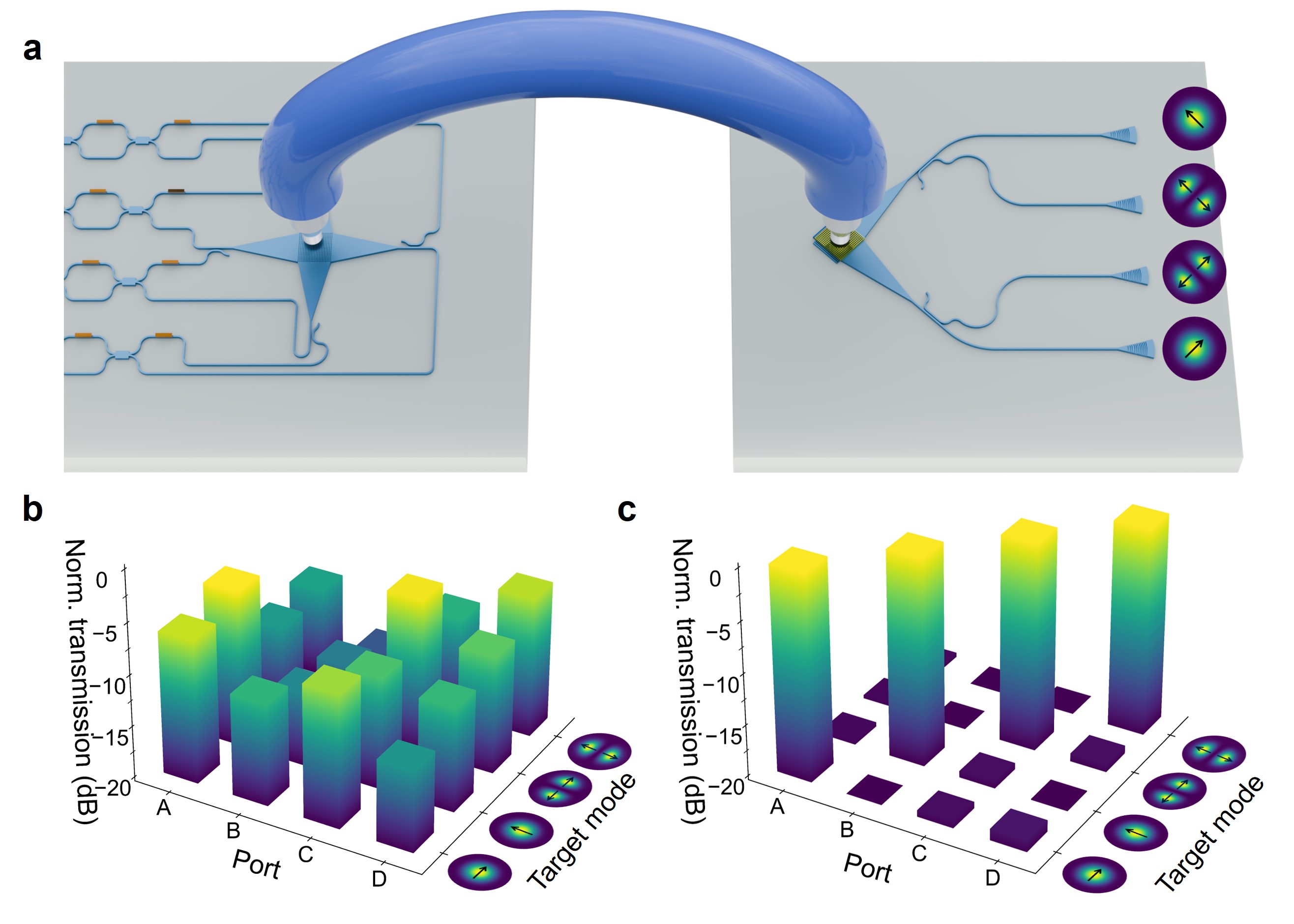}
    \caption{\textbf{Experimental setup and results for spatial and polarization distribution control in the attached optical fiber.} \textbf{a} Experimental setup to demonstrate the spatial and polarization control with two photonic chips. The two ends of optical fiber are attached to the two photonic chip via the multidimensional emitters. The multidimensional emitter on the right side is both spatial and polarization sensitive. Optical transmission power is measured to quantify the purity of spatial and polarization distribution at the fiber end face including \( \text{LP}_{01x} \), \( \text{LP}_{11a} \), \( \text{LP}_{11b} \), and \( \text{LP}_{01y} \). \textbf{b} Received initial optical power bar chart at the four output ports with unconfigured photonic mesh. The random distribution can vary with the fiber transmission matrix. The ports A to D represent the ports in the right photonics chip, designed to receive the \( \text{LP}_{01x} \), \( \text{LP}_{01y} \), \( \text{LP}_{11ax} \), and \( \text{LP}_{11by} \) modes, respectively. \textbf{c} Received optical power bar chart of the four spatial and polarization channels by accurately configure the photonic mesh to maximize the transmission for each physical channel.}
    \label{fig:5}
\end{figure*}

Until now, the spatial polarization distribution control has not been examined in our experiment, primarily because the infrared camera is polarization insensitive and unable to provide feedback on the polarization distribution of the imaged structure. As the integrated photonic processor can have full access to all the singular values in the optical fiber transmission matrix, spatial polarization control should also be allowed with appropriate feedback of the polarization status at the fiber end face. To demonstrate that in our experiment, a second photonic chip containing the emitter design shown in Figure \ref{fig:3}a is utilized at the output side of the optical fiber instead of infrared camera. The experimental setup is illustrated by Figure \ref{fig:5}a. The emitter with two injected waveguides is both polarization and mode sensitive and only support four spatial and polarization channels including \( \text{LP}_{\text{01x}} \), \( \text{LP}_{\text{01y}} \), \( \text{LP}_{\text{11ax}} \), and \( \text{LP}_{\text{11by}} \). One-to-one selective decoupling at the output fiber end face can be utilized to monitor the spatial and polarization distribution by measuring the optical transmission power from the the four on-chip waveguides.  Figure \ref{fig:5}b depicts the random transmission distribution bar chart with different initial structure launched by the left photonic processor. By maximize the transmission for each of the four spatial channels A-D as as shown in Figure \ref{fig:5}c, manipulation of the spatial and polarization distribution at the fiber end face can be achieved. The beam structure can be configured accurately as indicated by the low crosstalk measured over other spatial and polarization channels in Figure \ref{fig:5}c. For the spatial and polarization channels \( \text{LP}_{\text{01x}} \), \( \text{LP}_{\text{01y}} \), \( \text{LP}_{\text{11ax}} \), and \( \text{LP}_{\text{11by}} \), the measured inter-channel crosstalk values are -18.88\ \text{dB}, -19.37 \ \text{dB}, -19.68\ \text{dB}, and -19.56\ \text{dB}, respectively. Therefore, an accurate spatial and polarization control of the structure at the fiber end can also be realized by our method.

\section{Conclusion and outlook}\label{sec3}

To summarize, we demonstrated an efficient and adaptive wavefront shaping method for optical fibers by using a programmable integrated silicon photonic processor. Without any free space bulk optics or nanostructures attached to the fiber tips, precise manipulation of spatial and polarization distribution at the fiber end can be achieved by configuring the photonic chip electrically. Prior knowledge of the polarization rotation and inter-modal coupling in the optical fiber is not required. We further calculated the overlap integrals to validate the high beam quality generated from the photonic processor and from the fiber end face, respectively. In addition, we have introduced and validated a novel emitter structure to solve the concern of limited chip-to-fiber efficiency, using CMOS compatible fabrication process. Our proposed method offers several advantages, such as precise control, efficient emission, and adaptive reconfiguration. Notably, it offers scalability advantages as the photonic chip can be produced using commercial silicon photonic foundries. The optical fiber attachment process is also well-established, having been extensively utilized in various optical transceiver prototypes in data centers \cite{marchetti2019coupling}.

To serve other types of the structured light generation and manipulation with the reconfigurable photonic chip, a key component is the chip-to-fiber emitter, which will essentially determine the number of spatial and polarization channels, the beam structure integrity, and the emission efficiency. Real-time adaptive configuration of the photonic processor can be established in the future to handle the time-varying fiber transmission matrix due to variations such as fiber bending or twisting. This is feasible because the response time of thermal-optical phase shifters within the photonic mesh can be in the range of tens of microseconds or even less than a nanosecond utilizing electro-optical effects \cite{jacques2019optimization}. Enhanced phase tuning capabilities can be attained through lithium niobate integrated optics \cite{wang2018integrated}. Optimization of driving control electronic circuits and algorithms will  also be important in determining the implementation timeline for adaptive wavefront control. It is also worthwhile to mention that our proposed emitter structure can be applied for other future designs and improve their efficiency, which is relied on the optimized front-end-of-line fabrication process. Depending on the required dimensionality of the emitter, the integrated photonic mesh can also be easily scaled with more complex matrix operations for wavefront control. With the rapid growing maturity, other advanced photonic functions can be incorporated in the future including the heterogeneous integration with on-chip laser sources. The rapid development of the photonic integrated circuits show a great promise in achieving a portable and robust structured light generation and manipulation system in the future.

\section{Methods}\label{sec4}

\subsection{Photonic integrated circuits design and fabrication}
The photonic integrated circuits are fabricated in multi-project wafer (MPW) runs offered by imec and Applied Nanotools. The photonics chips are fabricated on a silicon-on-insulator (SOI) wafer with a crystalline silicon layer thickness of 220 nm. The buried-oxide (BOX) layer has a thickness is 2 $\mu$m. An etching depth of 70 nm is employed in the grating region to form the shallow etched holes and the low refractive index region. The strip waveguides are formed by a full etch process. An additional layer of polycrystalline silicon (Poly-Si) with a thickness of 160 nm is deposited on top for MPW runs offered by imec to enhance the directionality and emission efficiency. Top cladding of silicon dioxide and metallization are implemented. High-resistance titanium-tungsten alloy (TiW) is used for heater-based optical phase shifters. The photonic integrated processor is designed for operation around optical wavelength of 1550 nm. 

\subsection{Experimental setup and mesh configuration}
The graded-index few-mode fibers in our experiment are provided by OFS (Optical Fiber Solutions). An external tunable continuous-wave laser source (Santec TSL-770) is used with the multi-channel optical power meters (Santec MPM-210H and MPM-215). The integrated photonic processor is configured by a multichannel programmable power supply (TIME-TRANSFER T-MS128-12CV) and personal computer for the photonic mesh configuration. Far-field intensity distributions of the beam structure emitted from the chip and from the fiber end face are captured by an infrared camera (ARTCAM-991SWIR) with microscope objective. The photonic mesh is configured by particle-swarm algorithm according to the feedback of overlap integral between the captured structure and target field distribution.

\vspace{0.5cm}

\noindent \textbf{Data availability}

All the data that support the findings of this study are included in this article and its supplementary information. Any additional data are available from the corresponding author upon reasonable request.

\noindent \textbf{Supplementary information}

Additional details are attached in the supplementary materials.

\noindent \textbf{Author contributions}

W.Z., Z.C. and Y.T. conceived the idea. W.Z., Z.C., K.L., M.Z. and W.T. performed the experiment with the photonic integrated circuits. Z.C and W.Z designed the photonic integrated devices and circuits. W.Z., K.L. and H.C. contributed to printed circuit board design and control algorithms of the electronic system. W.Z and Y.T. wrote the manuscript with contributions from all authors. Y.T. supervised the project.

\noindent \textbf{Acknowledgements}:
Y.T. acknowledges the support from the National Natural Science Foundation of China (No. 62305277), Natural Science Foundation of Guangdong Province (No.2024A1515012438), Guangzhou Municipal Science and Technology Project (No.2023A03J0159, 2024A04J4234), and the Start-up fund from the Hong Kong University of Science and Technology (Guangzhou). The authors acknowledge the Novel IC Exploration (NICE) Facility of HKUST(GZ) for technical support. We thank imec and Applied Nanotools for photonic integrated circuits fabrication.

\bibliographystyle{naturemag}

\end{document}


\title[Article Title]{Supplementary Information for "Efficient and Adaptive Reconfiguration of Light Structure in Optical Fibers with Programmable Silicon Photonics"}


\author{Wu Zhou}
\author{Zengqi Chen}
\author{Kaihang Lu}
\author{Hao Chen}
\author{Mingyuan Zhang}
\author{Wenzhang Tian}
\author{Yeyu Tong}
\email{yeyutong@hkust-gz.edu.cn}
\affiliation{Microelectronic Thrust, The Hong Kong University of Science and Technology (Guangzhou), 511453, Guangzhou, Guangdong, PR China}

\maketitle
\tableofcontents
\listoffigures
\newpage

\section{Multidimensional chip-to-fiber emitter design}\label{ssec1}

\begin{figure}[b]
    \centering
    \includegraphics[width=0.8\linewidth]{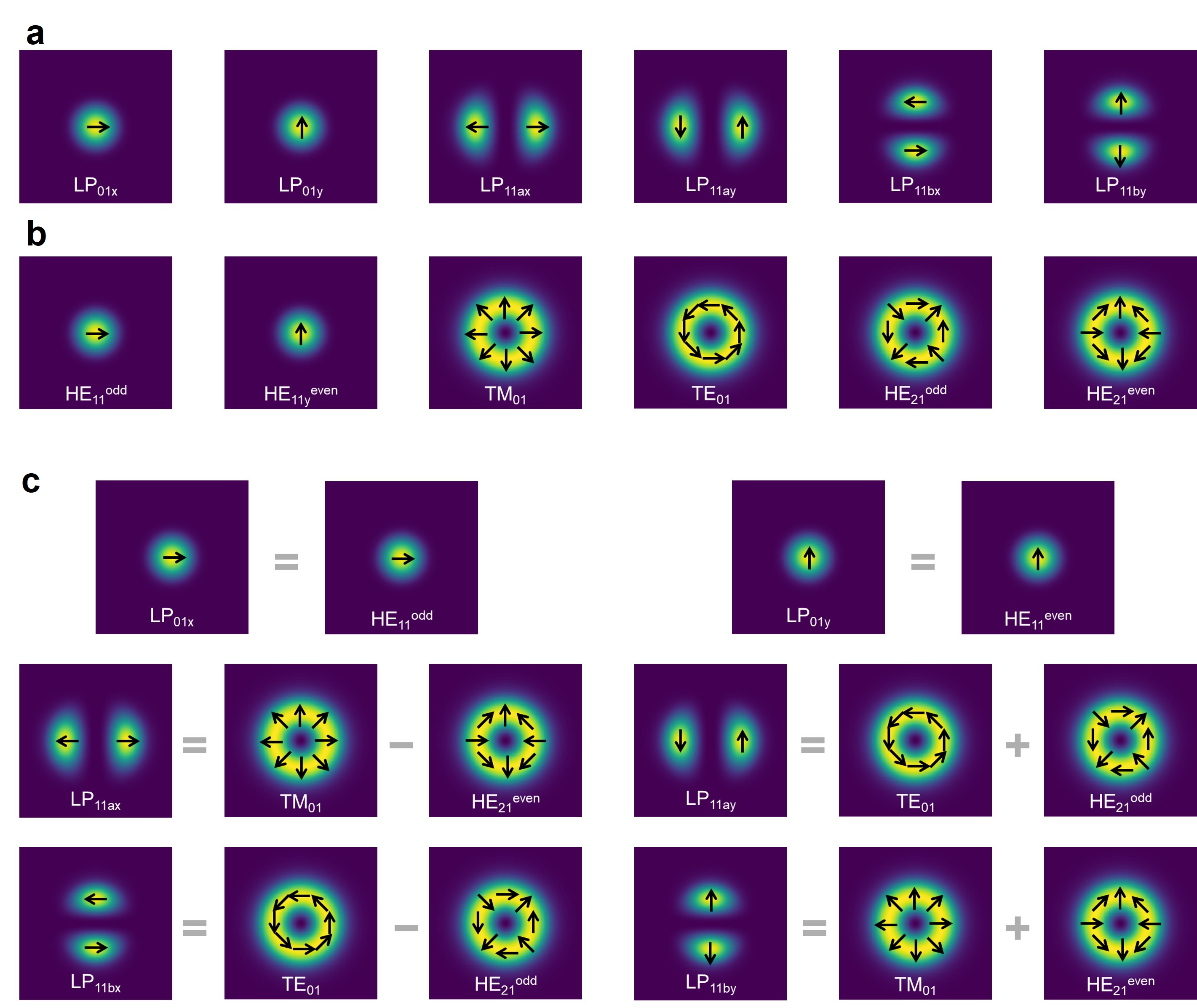}
    \caption{\textbf{Overview of six orthogonal spatial and polarization physical channels in a two-mode few-mode optical fiber.} linearly polarized (LP) mode is a description often used as a supervision of fiber eigenmodes. \textbf{a} LP mode field profile. \textbf{b} eigenmode field profile. \textbf{c} LP mode formed by the corresponding fiber eigenmodes. }
    \label{fig:s1}
\end{figure}

The graded-index two-mode few-mode optical fiber in our experiment is provided by OFS (Optical Fiber Solutions), with a mode field diameter of around 11.0 $\mu$m for the two linearly polarized (LP) mode groups, \( \text{LP}_{01} \) and \( \text{LP}_{11} \). Totally there exists six orthogonal spatial and polarization channels including \( \text{LP}_{\text{01x}} \), \( \text{LP}_{\text{01y}} \), \( \text{LP}_{\text{11ax}} \), \( \text{LP}_{\text{11ay}} \), \( \text{LP}_{\text{11bx}} \) and \( \text{LP}_{\text{11by}} \), as depicted by Figure \ref{fig:s1}a. Another often used description is the fiber eigenmodes including \( \text{HE}_{11}^{\text{odd}} \), \( \text{HE}_{11}^{\text{even}} \), \( \text{TM}_{01} \), \( \text{TE}_{01} \), \( \text{HE}_{21}^{\text{odd}} \), \( \text{HE}_{21}^{\text{even}} \), as depicted by Figure \ref{fig:s1}b. The correspondence between the fiber eigenmodes and the LP modes are illustrated by Figure \ref{fig:s1}c, which shows that both can be utilized to describe the full set of six orthogonal channels supported by the two-mode few-mode optical fiber.

\begin{figure}[t]
    \centering
    \includegraphics[width=0.8\linewidth]{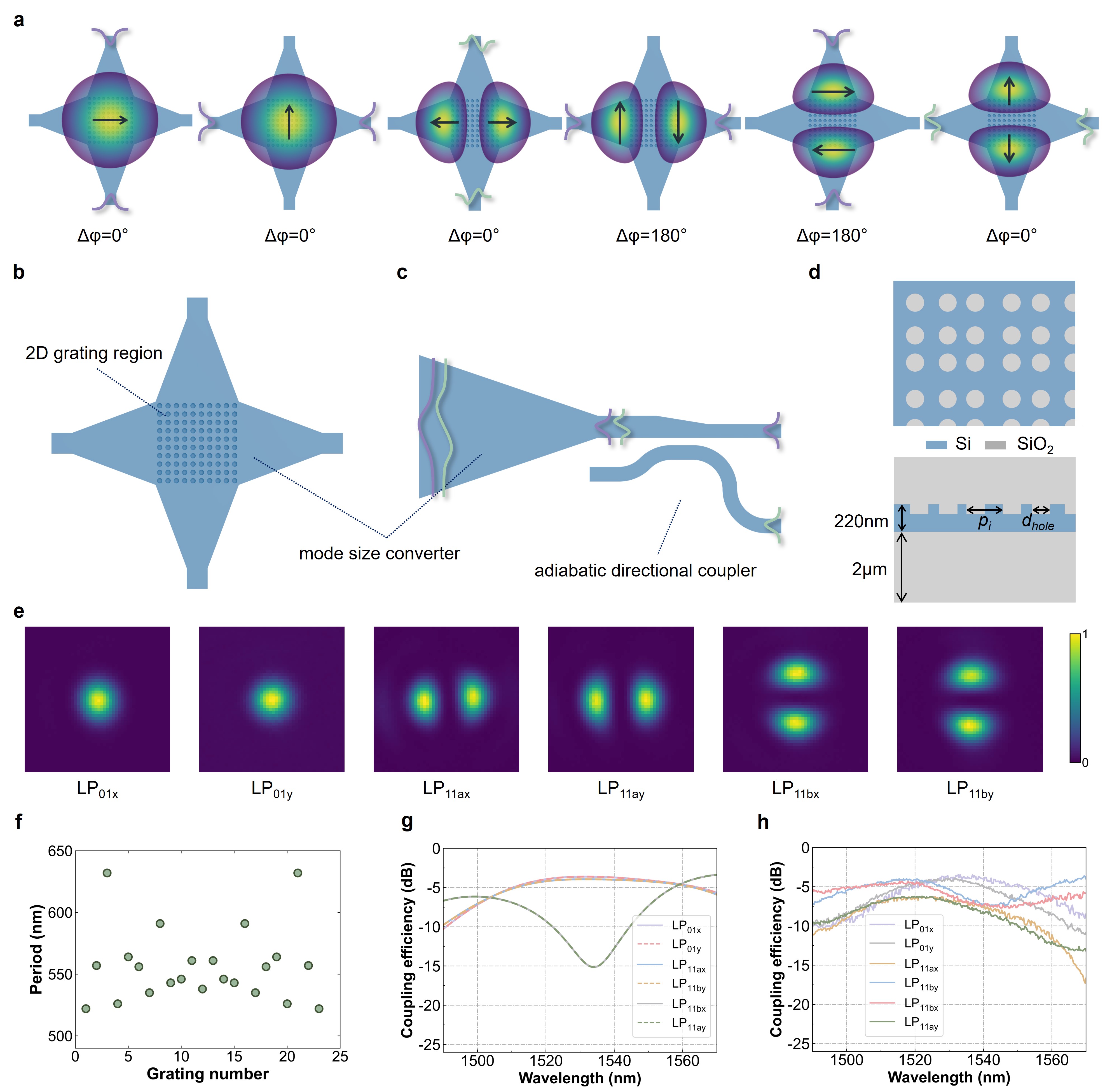}
    \caption{\textbf{Multidimensional chip-to-fiber emitter design and working principle.} \textbf{a} selective mode excitation conditions for the proposed multidimensional chip-to-fiber emitter. \textbf{b} topic view of the emitter including 2D grating and mode size converter. \textbf{c} zoom-in view of the mode size converter and adiabatic directional coupler. \textbf{d} zoom-in view of the 2D grating region with shallowly etched holes and cross sectional view.  \textbf{e} diffracted beam structure form the emitter captured by an infrared camera under different selective launching conditions. \textbf{f} grating periods of the shallowly etched holes. \textbf{g} coupling efficiency for spectra for various fiber modes obtained in FDTD simulation. \textbf{h} experimental coupling efficiency spectra for various fiber modes.}
    \label{fig:s2}
\end{figure}

The multidimensional chip-to-fiber emitter is designed to support the full set of six spatial and polarization modes exist in a two-mode few-mode optical fiber. Selective excitation of each fiber LP mode is illustrated by Figure \ref{fig:s2}a, where each fiber mode is launched by two counter-propagating transverse electric (TE) modes with appropriate phase delay. When the relative phase delay is equal to zero, \( \text{LP}_{\text{01x}} \), \( \text{LP}_{\text{01y}} \), \( \text{LP}_{\text{11ax}} \), and \( \text{LP}_{\text{11by}} \) can be excited selectively by two counter-propagating \( \text{TE}_{0} \) or \( \text{TE}_{1} \) mode. When the relative phase difference is equal to $\pi$, \( \text{LP}_{\text{11ay}} \) and \( \text{LP}_{\text{11bx}} \) can be excited. All the injected eight TE modes share the same 2D grating region which is centrosymmetric. The simultaneous launching of \( \text{TE}_{0} \) and \( \text{TE}_{1} \) mode is allowed as their effective refractive indices are almost the same in the grating region with a width of more than 13 $\mu$m.  Figure \ref{fig:s2}b-\ref{fig:s2}c shows the top view of the schematic including 2D grating region with 70-nm shallowly etched holes, mode size converter based on linear adiabatic taper \cite{wang2013silicon, ding2013chip}, and asymmetrical adiabatic directional coupler based on evanescent field coupling. The cross-sectional view of the emitter is shown in Figure \ref{fig:s2}d, which is based on the silicon-on-insulator (SOI) platform with 220-nm thick crystalline silicon and 2-$\mu$m buried oxide. The grating periods and hole diameter in our design are optimized by optimization algorithm and finite-difference time-domain (FDTD) simulation. Detailed design method can be found in \cite{zhou2024ultra}. The measured beam structures directly from the emitter for the six orthogonal spatial and polarization channels are shown in Figure \ref{fig:s2}e. Figure \ref{fig:s2}f and \ref{fig:s2}g present the design parameters and simulation coupling efficiencies for various fiber modes, respectively. The measured experimental coupling efficiencies for various fiber modes are depicted in Figure \ref{fig:s2}h.

\section{Polycrystalline silicon overlay enhanced emitter}\label{ssec2}

\begin{figure}[t]
    \centering
    \includegraphics[width=1.0\linewidth]{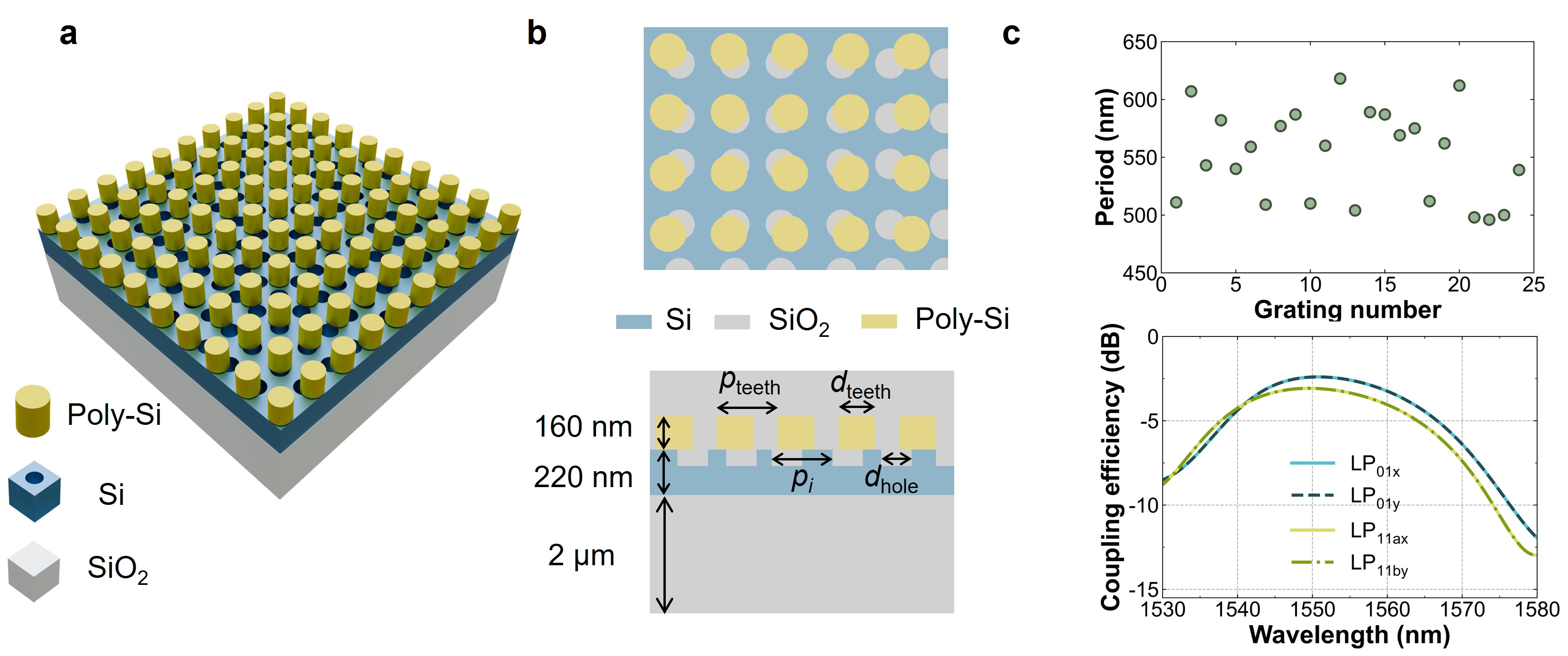}
    \caption{\textbf{Polycrystalline silicon (poly-Si) overlay enhanced emitter design.} \textbf{a} schematic design of the 2D grating with shifted poly-Si overlay. \textbf{b} Zoom-in view and cross sectional view of the grating region. The poly-Si disks and the shallowly etched holes are staggered intentionally. \textbf{c} Optimized design parameters and simulated coupling efficiencies for various LP modes in the optical fiber.}
    \label{fig:s3}
\end{figure}

While the previous multidimensional chip-to-fiber emitter shows satisfying coupling efficiencies for various fiber modes, it is desired to further increase its efficiency performance while using complementary metal-oxide semiconductor (CMOS) compatible process and optical lithography. Figure \ref{fig:s3}a depicts the new emitter proposed in this work with polycrystalline silicon (poly-Si) overlay. Such fabrication process has been standardized in the commercial silicon photonic multi-project wafer (MPW) service offered by photonic foundries. The corresponding top-view and cross-sectional view of the multidimensional emitter are illustrated by Figure \ref{fig:s3}b. The 160-nm thick poly-Si disks are intentionally shifted over the 70-nm shallowly etched holes, to allow precise control of the effective refractive index in the 2D grating region. By doing so, the grating strength, which is defined as the diffracted energy percentage per unit grating length, can be engineered to optimize its field overlap integral with the field distribution supported in the few-mode optical fiber. In order to improve the coupling efficiency and mitigate second-order Bragg reflections that arise from a perfectly vertical coupling configuration, key parameters such as the nanohole diameter ($d_{\text{hole}}$), the nanohole periodicity ($p_i$), the poly-Si disk diameter ($d_{\text{teeth}}$), and the poly-Si disk periodicity ($p_{\text{teeth}}$) (illustrated in Figure~\ref{fig:s3}b) were optimized using genetic algorithms in combination with FDTD simulations. Figure~\ref{fig:s3}c presents the optimized parameter distributions for the nanohole periods ($p_i$) and the coupling efficiency for each mode in the few-mode fiber. Specifically, the nanohole diameter ($d_{\text{hole}}$) is 325~nm, the poly-Si disk diameter ($d_{\text{teeth}}$) is 300~nm, and the poly-Si disk period ($p_{\text{teeth}}$) is 536~nm, respectively. The use of the poly-Si overlay leads to a significant enhancement in coupling efficiency across all the spatial and polarization modes supported in the two-mode few-mode optical fiber. It is also worthwhile to mention that, the emitter feeding from two perpendicular directions can support selective launching of four fiber LP modes by using a single TE mode on chip, while all the six spatial and polarization modes can be excited by injecting from four perpendicular directions in the future. The poly-Si overlay enhanced emitter here will also be used later in our experiment at the other output end face of the optical fiber, to selectively decouple various spatial and polarization modes and quantify the inter-channel crosstalk, which cannot be realized via a polarization-insensitive infrared camera.

\section{Mathematics of the structured light generation system via photonic integrated processor}\label{ssec3}

\begin{figure}[t]
    \centering
    \includegraphics[width=1.0\linewidth]{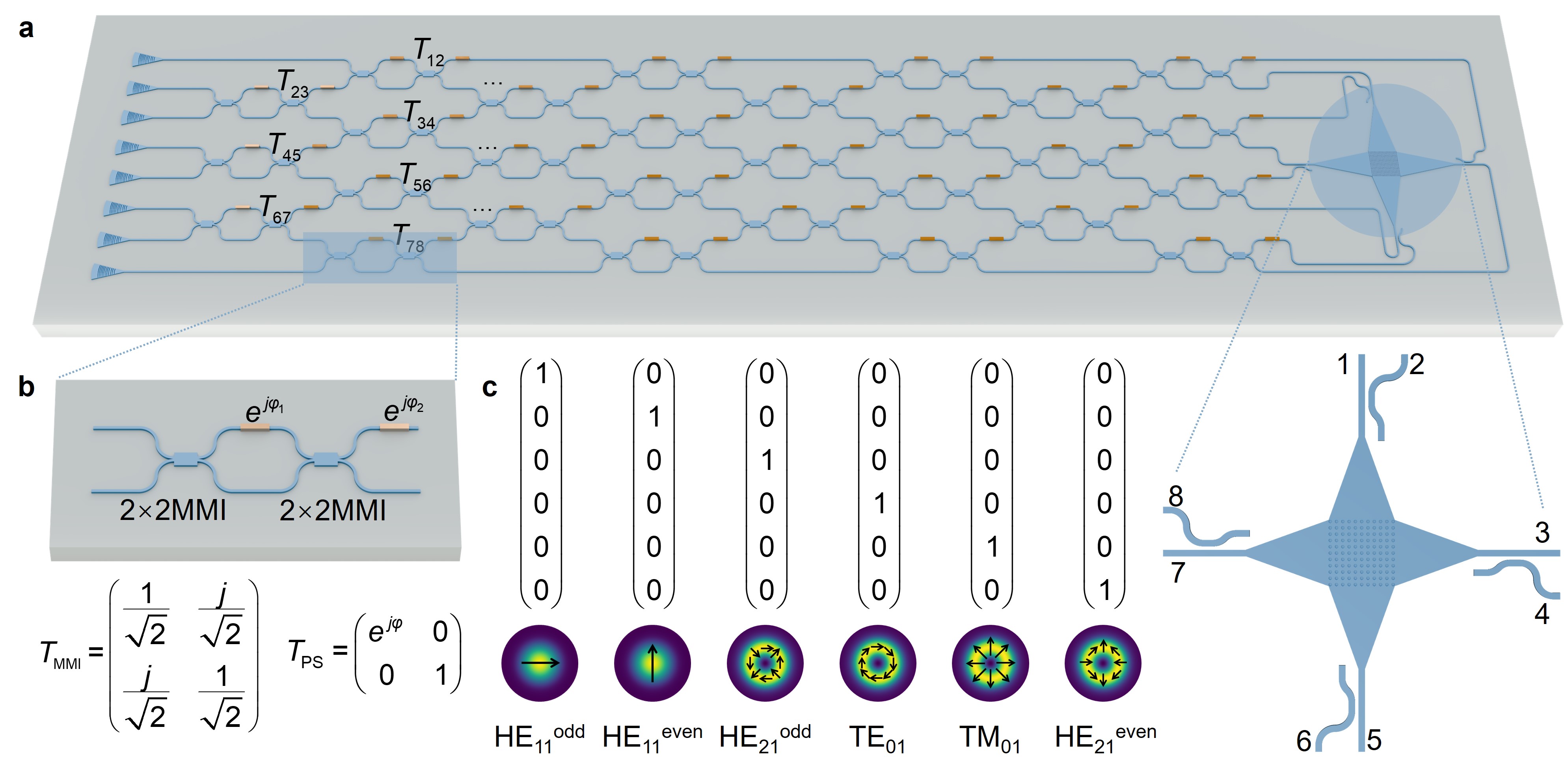}
    \caption{\textbf{Overview of the programmable photonic integrated processor.} \textbf{a} Three-dimensional schematic of the programmable photonic integrated processor, comprising a mesh of MZI networks and a multidimensional emitter. \textbf{b} A single MZI unit within the processor, capable of performing a unitary transformation represented by a \( U(2) \) matrix. \textbf{c} The multidimensional emitter with eight input ports, enabling independent selective excitation of a complete set of eigenmodes in a two-mode fiber. }
    \label{fig:s4}
\end{figure}

The programmable photonic integrated processor employs an 8×8 unitary optical mesh $U_{8}$ to control the amplitude and phase of light at the eight input ports of the multimode grating emitters. The unitary optical mesh consists of a rectangular array\cite{clements2016optimal} of 28 tunable 2×2 Mach-Zehnder interferometers. Figure \ref{fig:s4}b illustrates a schematic of the tunable 2×2 Mach-Zehnder interferometer unit, consisting of two 50:50 multimode interferometers(MMIs) and two heater-based optical phase shifters(PSs). The scattering matrix of each Mach-Zehnder interferometer can be derived by sequentially multiplying the scattering matrices of the PSs and the MMIs, as shown in Equation \ref{mzi_unit_matrix}. The simplified result, derived using Euler's formula, is presented in Equation \ref{mzi_unit_matrix_simplify}. This result clearly demonstrates that the system can perform a unitary transformation $U_{2}$, as verified by Equation \ref{unitary_matrix_condition}. As illustrated in Figure \ref{fig:s2}a, in a rectangular array network, each Mach-Zehnder interferometer can be cascaded topologically to form an 8×8 unitary transformation matrix. The resulting matrix is presented in Equation \ref{unitary_matrix}, where \( S \) defines the specific ordered sequence of unitary transformations executed by the BSs. The expression \( T_{m,n} \), given in Equation \ref{sub_unitary_matrix}, describes the two-dimensional subspace modified by each BS within the $U_{8}$ transformation.

\begin{equation}
S = 
\begin{pmatrix}
e^{j \varphi_1} & 0 \\
0 & 1
\end{pmatrix}
\begin{pmatrix}
\frac{1}{\sqrt{2}} & \frac{j}{\sqrt{2}} \\
\frac{j}{\sqrt{2}} & \frac{1}{\sqrt{2}}
\end{pmatrix}
\begin{pmatrix}
e^{j \varphi_2} & 0 \\
0 & 1
\end{pmatrix}
\begin{pmatrix}
\frac{1}{\sqrt{2}} & \frac{j}{\sqrt{2}} \\
\frac{j}{\sqrt{2}} & \frac{1}{\sqrt{2}}
\end{pmatrix}
\label{mzi_unit_matrix}
\end{equation}

\begin{equation}
S = j e^{j\frac{\varphi_2}{2}}
\begin{pmatrix}
e^{j\varphi_1} \sin{\frac{\varphi_2}{2}} & e^{j\varphi_1} \cos{\frac{\varphi_2}{2}} \\
\cos{\frac{\varphi_2}{2}} & -\sin{\frac{\varphi_2}{2}}
\end{pmatrix}
\label{mzi_unit_matrix_simplify}
\end{equation}

\begin{equation}
SS^\dagger = S^\dagger S = I
\label{unitary_matrix_condition}
\end{equation}

\begin{equation}
U = \prod_{(m,n) \in S} T_{m,n}
\label{unitary_matrix}
\end{equation}

\begin{equation}
T_{m,n} = \begin{pmatrix}
1 & 0 & \cdots & \cdots & \cdots & \cdots & 0 \\
0 & 1 & \cdots & \cdots & \cdots & \cdots & 0 \\
\vdots & \vdots & \vdots & \vdots & \vdots & \vdots & \vdots \\
0 & 0 & \cdots & e^{j \varphi_1} \sin{\frac{\varphi_2}{2}} & e^{j \varphi_1} \cos{\frac{\varphi_2}{2}} & \cdots & 0 \\
0 & 0 & \cdots & \cos{\frac{\varphi_2}{2}} & -\sin{\frac{\varphi_2}{2}} & \cdots & 0 \\
\vdots & \vdots & \vdots & \vdots & \vdots & \vdots & \vdots \\
0 & 0 & \cdots & \cdots & \cdots & \cdots & 1
\end{pmatrix}
\label{sub_unitary_matrix}
\end{equation}

In a two-mode fiber, there are six eigenmodes, specifically: \( \text{HE}_{11}^{\text{odd}} \), \( \text{HE}_{11}^{\text{even}} \), \( \text{HE}_{21}^{\text{odd}} \), \( \text{TE}_{01} \), \( \text{TM}_{01} \), and \( \text{HE}_{21}^{\text{even}} \). The transmission matrix of the two-mode FMF is a \( 6 \times 6 \) unitary matrix $U_{6}$. A unitary matrix possesses the property of invertibility, with its inverse being equivalent to its conjugate transpose $U_{6}^{*}$. To derive the matrix operations for structured light generation, as shown in Figure \ref{fig:s4}c, we define each eigenmode in the two-mode fiber as a \( 6 \times 1 \) unit vector. These vectors are mutually orthogonal, forming a set of orthogonal basis vectors in the two-mode fiber. These eigenmodes form an orthonormal basis set, which allows us to describe any light field within the fiber as a linear superposition of these eigenmodes. After going through the optical mesh with reconfigurable unitary transformation $U_{8}$, an efficient multidimensional chip-to-fiber emitter as shown in S1 is employed before the few-mode fiber (FMF). The 2D grating coupler is able to convert each of the eight on-chip single-mode channels (1-8) into the six eigenmodes of the two-mode FMF. Therefore, under the ideal condition where losses are neglected, the transmission matrix of the multidimensional emitter can be represented by a \( 6 \times 8 \) matrix \( M_{6 \times 8} \). By calculating the eigenmode components excited by each port after passing through the grating coupler, it can be easily shown that \( M_{6 \times 8} \) follows the form described in Equation \ref{GC_transmission_matrix}. It can be proven through Equation \ref{GC_matrix_orthogonality} that matrix \( M_{6 \times 8} \) is indeed a semi-unitary matrix.

\begin{equation}
M_{6 \times 8} = \begin{pmatrix}
\frac{\sqrt{2}}{2} & 0 & 0 & 0 & \frac{\sqrt{2}}{2} & 0 & 0 & 0 \\
0 & 0 & \frac{\sqrt{2}}{2} & 0 & 0 & 0 & \frac{\sqrt{2}}{2} & 0 \\
\frac{1}{2} & 0 & \frac{1}{2} & 0 & -\frac{1}{2} & 0 & -\frac{1}{2} & 0 \\
\frac{1}{2} & 0 & -\frac{1}{2} & 0 & -\frac{1}{2} & 0 & \frac{1}{2} & 0 \\
0 & -\frac{1}{2} & 0 & \frac{1}{2} & 0 & -\frac{1}{2} & 0 & \frac{1}{2} \\
0 & \frac{1}{2} & 0 & \frac{1}{2} & 0 & \frac{1}{2} & 0 & \frac{1}{2} \\
\end{pmatrix}
\label{GC_transmission_matrix}
\end{equation}

\begin{equation}
M_{6 \times 8} \cdot M_{6 \times 8}^{\dagger} = I
\label{GC_matrix_orthogonality}
\end{equation}

When an arbitrary input optical signal \( x_{\text{in}} \) is launched from input single mode grating couplers, the speckle pattern after passing through the on-chip photonic mesh and optical fiber, can be described by Equation \ref{mode_transfermation_matrix_1}. Each element of \( y_{\text{out}} \) is a complex value, representing the amplitude and phase of the corresponding components of the eigenmodes. In our experiment, the input signal is launched by one of input single-mode grating couplers using an external laser source. The product of the four matrices, \( M_{6 \times 8}^{\dagger} \), \( U_{6}^{\dagger} \), \( M_{6 \times 8} \), and \( U_{8} \), can form a unitary matrix, as demonstrated by Equation \ref{u8decom}. Consequently, it can be represented by the on-chip optical mesh in our design. By substituting this product into Equation \ref{mode_transfermation_matrix_1}, we can derive Equation \ref{mode_transfermation_matrix_2}, which demonstrates that the on-chip reconfigurable optical mesh can always invert the unitary transmission matrix of the two-mode few-mode optical fiber. As the \(M_{6 \times 8} \cdot U_{8}\) in Equation \ref{mode_transfermation_matrix_2} is a semi-unitary matrix, arbitrary rotation of the input vector with a constant modulus can thus be realized, which can generate arbitrary guided light structure at the fiber end by superposition of all the possible eigenmodes as shown in Equation \ref{modes} and \ref{intensity}.

\begin{equation}
y_{\text{out}} = U_{6} \cdot M_{6 \times 8} \cdot U_{8} \cdot x_{\text{in}}
\label{mode_transfermation_matrix_1}
\end{equation}

\begin{equation}
    M_{6 \times 8}^{\dagger} U_{6}^{\dagger} M_{6 \times 8} U_{8} \cdot \left( M_{6 \times 8}^{\dagger} U_{6}^{\dagger} M_{6 \times 8} U_{8} \right)^{\dagger} = \left( M_{6 \times 8}^{\dagger} U_{6}^{\dagger} M_{6 \times 8} U_{8} \right)^{\dagger} \cdot M_{6 \times 8}^{\dagger} U_{6}^{\dagger} M_{6 \times 8} U_{8} = E
    \label{u8decom}
\end{equation}

\begin{equation}
y_{\text{out}} = U_{6} \cdot M_{6 \times 8} \cdot \left( M_{6 \times 8}^{\dagger} U_{6}^{\dagger} M_{6 \times 8} U_{8} \right) \cdot x_{\text{in}} = M_{6 \times 8} \cdot U_{8} \cdot x_{\text{in}}
\label{mode_transfermation_matrix_2}
\end{equation}

\begin{equation}
y_{\text{out}} = \begin{pmatrix}
A_1 e^{j\varphi_1} \\
A_2 e^{j\varphi_2} \\
A_3 e^{j\varphi_3} \\
A_4 e^{j\varphi_4} \\
A_5 e^{j\varphi_5} \\
A_6 e^{j\varphi_6} \\
\end{pmatrix}
= \sum_{i=1}^{6} A_i e^{j\varphi_i} \mathbf{F}_i(x,y)
\label{modes}
\end{equation}

\begin{equation}
I = \left| \sum_i \left( A_i e^{j\varphi_i} \right) \mathbf{F}_i(x,y) \right|^2
\label{intensity}
\end{equation}

\section{Acquisition and characterization of beam structures}\label{ssec4}

In this experiment, a 10× objective lens and an infrared camera were utilized to capture the structured light emitted from both the emitter and the fiber end face. To analyze the light distribution, the phase shifters inside the photonic mesh were tuned to compute the similarity between the captured light patterns under different power configurations and a desired target pattern. Particle Swarm Optimization (PSO) \cite{kennedy1995particle} was employed to determine the optimal voltage configuration that maximized this similarity, effectively training the photonic processor to reproduce arbitrary structured light patterns. It is thus necessary to compute the structured light distribution aligned with the few-mode fiber utilized in the setup. Furthermore, an accurate method for quantifying the similarity between the light pattern captured by the infrared camera and the target pattern was essential for precise comparative analysis.

The target light pattern was obtained by numerically solving the Helmholtz equation \cite{feit1979calculation}, as shown in Equation \ref{Helmholtz_equation}, where $E$ represents the electric field distribution, $k = \frac{2\pi}{\lambda}$ is the wavenumber, and $n$ is the refractive index. For the few-mode fiber system, this equation was reformulated in cylindrical coordinates (Equation \ref{Helmholtz_equation_Cylindrical_coordinate}), with parameters specific to the fiber used in the experiment. By incorporating fiber parameters such as the core radius $r_{\text{core}}$, core refractive index $n_{\text{core}}$, cladding refractive index $n_{\text{cladding}}$, and operating wavelength $\lambda$, the numerically computed structured light distribution was aligned with the experimental conditions.

\begin{equation}
\nabla^2 E + k^2 n^2 E = 0
\label{Helmholtz_equation}
\end{equation}

\begin{equation}
\begin{aligned}
\frac{1}{r}\frac{d}{dr}\left( r\frac{dE(r)}{dr} \right) + \left( \left( \frac{2\pi}{\lambda} \right)^2 n^2(r) - \beta^2 \right)E(r) = 0, \\
n(r) = 
\begin{cases} 
n_{\text{core}}, & r \leq r_{\text{core}} \\
n_{\text{cladding}}, & r > r_{\text{core}}
\end{cases}
\end{aligned}
\label{Helmholtz_equation_Cylindrical_coordinate}
\end{equation}

To quantify the similarity between the structured light pattern generated by the photonic processor and the target pattern, the overlap integral is employed. The overlap integral is a widely used tool in optics and photonics to measure the degree of spatial overlap between two electromagnetic field distributions. It provides a robust metric for comparing the spatial modes of light, making it an ideal choice for evaluating the similarity of the beam patterns in this experiment.

\begin{figure}[t]
    \centering
    \includegraphics[width=1.0\linewidth]{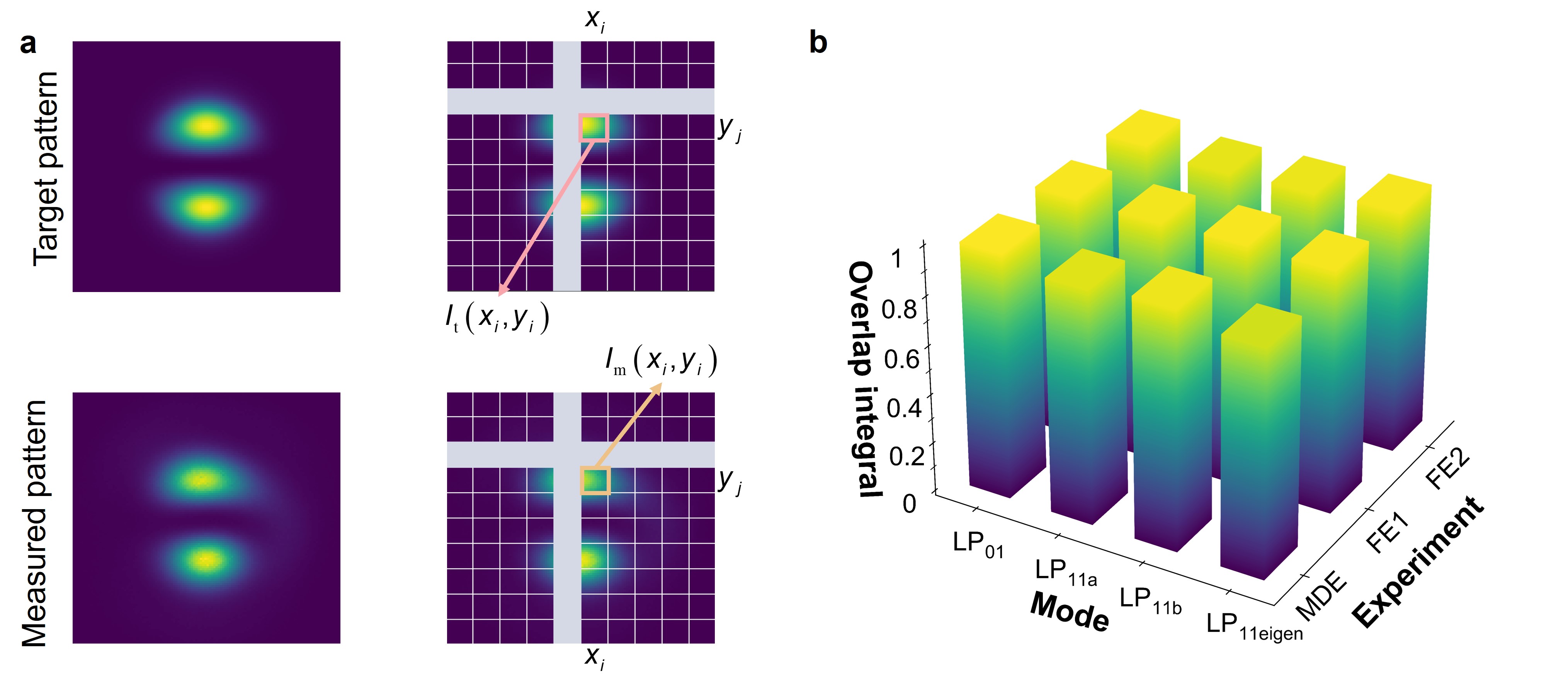}
    \caption{\textbf{Characterization of beam structures through overlap integrals.} \textbf{a} Principle of intensity-based overlap integrals, estimated by accumulating discrete points. \textbf{b} Overlap integrals between the generated structured beams and the target beams under three different experimental conditions. The MDF represents the results for four structured beams generated by the multidimensional emitter, while FE1 and FE2 correspond to the experimental results of the structured beams generated at the fiber output under different fiber stress conditions. }
    \label{fig:s5}
\end{figure}

The overlap integral is defined as the normalized inner product of two field distributions, the measured field \( E_{\text{m}}(x, y) \) and the target field \( E_{\text{t}}(x, y) \). Mathematically, the overlap integral is expressed as:

\begin{equation}
O_i = \frac{\left| \iint E_{\text{m}}(x, y) E_{\text{t}}^*(x, y) \, dx \, dy \right|^2}{\iint |E_{\text{m}}(x, y)|^2 \, dx \, dy \iint |E_{\text{t}}(x, y)|^2 \, dx \, dy}
\label{overlap_integral}
\end{equation}

In this equation, \( E_{\text{m}}(x, y) \) represents the electric field distribution of the light pattern captured by the infrared camera, while \( E_{\text{t}}(x, y) \) is the electric field distribution of the target pattern. The symbol \( ^* \) denotes the complex conjugate of the target field. The numerator in Equation \ref{overlap_integral} calculates the squared modulus of the inner product (or overlap) between the two field distributions, providing a measure of how well the two patterns align spatially. The denominator normalizes this value by the total energy of each field, ensuring that the similarity measure is independent of the absolute intensity of the fields.

However, in our experiment, the measurement is constrained by the nature of the infrared camera, which only captures the intensity distribution at discrete pixel points. As a result, the exact electromagnetic field distributions \( E(x, y) \) cannot be directly retrieved. Instead, we rely on the measured intensity distribution, \( I_{\text{m}}(x, y) = |E_{\text{m}}(x, y)|^2 \), which gives us the real-valued intensity information at each pixel. Consequently, we modify the method for calculating the overlap integral \( O_i \), replacing the complex field distributions with the available intensity data.

As depicted in Figure \ref{fig:s5}a, the real-valued overlap integral is derived by summing the pixel-wise products of the intensity distributions between the measured and target patterns \cite{lu2014adaptive}:

\begin{equation}
O_{\text{real}} = \frac{\sum_{i,j} \left(I_{\text{m}}(x_i, y_j) - \overline{I_{\text{m}}}\right)\left(I_{\text{t}}(x_i, y_j) - \overline{I_{\text{t}}}\right)}{\sqrt{\sum_{i,j} \left(I_{\text{m}}(x_i, y_j) - \overline{I_{\text{m}}}\right)^2 \sum_{i,j} \left(I_{\text{t}}(x_i, y_j) - \overline{I_{\text{t}}}\right)^2}}
\label{real_overlap_integral_avg}
\end{equation}

Here, \( I_{\text{m}}(x_i, y_j) \) represents the intensity recorded at the pixel \( (x_i, y_j) \) for the measured pattern, and \( I_{\text{t}}(x_i, y_j) \) represents the intensity at the same pixel for the target pattern. The numerator calculates the sum of the pixel-wise products of the two intensity distributions, which serves as a measure of their similarity. The denominator normalizes the result by the total intensity of both patterns, ensuring that the overlap measure is independent of the overall intensity scaling. The pixel-based summation method provides a practical and reliable means to evaluate the similarity between the experimentally obtained beam pattern and the target pattern.

Figure \ref{fig:s5}b presents the overlap integrals between the structured beams and target beams generated in the three experimental groups described in the main text. Among them, MDE corresponds to one experiment where the structured beams are generated by the multidimensional emitter. FE1 and FE2 refer to the other two experiments, in which the structured beams are generated at the fiber output under different fiber stress conditions. All structured beams show a high beam quality, achieving overlap integrals around 0.92 or above.

\bibliographystyle{naturemag}